\documentclass{article}
\usepackage{amsmath,times,bm,bbm,amssymb}
\usepackage{natbib}

\begin{document}

\title{Bayesian model choice via mixture distributions with application to epidemics and population process models}
\date{}
\author{P. D. O'Neill, T. Kypraios\\ School of Mathematical Sciences, University of Nottingham, UK\\
{\em philip.oneill@nottingham.ac.uk}\\ {\em theodore.kypraios@nottingham.ac.uk}}

\maketitle

\begin{abstract}
We describe a new method for evaluating Bayes factors. The key idea is to introduce
a hypermodel in which the competing models are components of a mixture distribution.
Inference for the mixing probabilities then yields estimates of the Bayes factors.
Our motivation is the setting where the observed data are a partially observed
realisation of a stochastic population process, although the methods have far wider
applicability. The methods allow for missing data and for parameters to be shared
between models. Illustrative examples including epidemics, population processes and
regression models are given, showing that the methods are competitive compared to
existing approaches.
\end{abstract}
{\em Keywords: Bayes factors; Data imputation; Epidemic models; Markov chain Monte Carlo methods; Model choice}

\section{Introduction}
\label{s:intro}
This paper describes a new method for evaluating Bayes factors. The method
itself is quite general, and can be applied to data consisting of independent and
identically distributed observations, or to data arising from more complex scenarios.
Our motivation comes from the setting of partially observed population processes in
which missing data imputation is typically required to facilitate inference.
We start by describing the motivating problem before considering the general context.

Consider an observed sequence of event times, each event being of the same type,
and suppose we wish to assess whether a homogeneous Poisson
process or an alternative non-homogeneous Poisson process best fits the
observations. Alternatively, suppose we have case-detection times in an outbreak
of infectious disease, and wish to know which of two possible
Susceptible-Infective-Removed disease transmission models is most plausible
as a model for how the data were generated, assuming that removals correspond
to case-detections. Both of these examples are special cases of a generic situation in which we wish
to assess which of a number of proposed point-process models best fits the data to hand. In
the first example there is one type of event, and all events are observed. In the
second example there are two event types (infections and removals) but only
the latter are observed. In both examples two models are compared, but in general
we may have more models of interest.

In a Bayesian framework, questions of model choice can be addressed using Bayes
factors, which quantify the relative likelihood of any two models given the data
and within-model prior distributions. Bayes factors can suffer from
two practical drawbacks, namely (i) they can be difficult to compute, and (ii)
they can be highly sensitive to the choice of within-model prior distributions,
and in particular apparently natural choices can give misleading results.
Here our focus is towards addressing the first difficulty, but in respect of the
second we briefly remark that alternative methods of Bayesian model assessment
have their own difficulties in the setting we consider. For example, neither
the Deviance Information Criterion nor Bayesian Information Criterion
appear entirely natural for settings where the data are typically far from
being independent observations, as is the case when the data are realisations
of a stochastic process. For problems involving missing data, such as the epidemic
example above, it is not even clear how suitable information criteria should best
bbe defined (\cite{Celeux06} give nine candidates, for instance). Finally, methods
involving a comparison between the observed data and what the fitted model would
predict typically involve a subjective judgement as to precisely what should be
compared, and how.

In all but the simplest cases, Bayes factors must be evaluated numerically. In principle, this can be achieved via
reversible jump Markov chain Monte Carlo methods \citep{Gr95}. To be precise, consider two models
$M_1$ and $M_2$ with parameters $\theta_1$ and $\theta_2$, respectively, where $\theta_j
\in \Theta_j$.  Define $k \in \left\{ 1, 2 \right\}$ to be a model indicator which specifies
the model under consideration. Reversible jump methods proceed by defining a Markov chain with state space
$\left\{ 1 \right\} \times \Theta_1 \cup \left\{ 2 \right\} \times \Theta_2$ such that the proportion of
time for which $k=j$ converges to the posterior model probability $P(M_j | x)$, where $x$ denotes the observed data.
Given model prior probabilities $P(M_j)$, the Bayes factor in favour of model 1 is given by the
expression $P(M_2)P(M_1|x)/P(M_1)P(M_2|x)$, which can be estimated from the Markov chain output.
The main practical challenge in implementing reversible jump algorithms is constructing efficient
between-model proposal distributions, i.e. defining how the Markov chain jumps between
the different components of the union of model parameter spaces. Although there have
been theoretical advances which address this issue \citep{Brooks_etal03}, for many
problems it remains a case of trial and error.

In this paper we propose a new method for evaluating Bayes factors which goes some way to
removing the implementation challenges of reversible jump methods. The key idea is to consider a hypermodel
which is itself a mixture model whose components are the two or more competing models
of interest. A Markov chain Monte Carlo algorithm can then be defined on the product space of all model
parameters and mixture probabilities. Our key result, Lemma 1, shows that the Bayes factors for the models can be
expressed in terms of the posterior means of the mixture probabilities, and thus
estimated from the Markov chain output. Our methods allow the incorporation of missing data, and for
model parameters to be shared between models.

Before proceeding to the details, we consider the general context. First, defining a Markov chain
on a product (rather than union) of model-parameter spaces is the approach pioneered
by \cite{CarChib95}, and further developed to more general settings (\cite{GrOHagan98},
\cite{Dell_etal02}, \cite{God01}). This approach, as for
reversible jump methods, involves defining a probability distribution over the set of possible models,
and introduces a parameter which indicates which model is chosen. In our setting the possible
models are combined into one mixture model.
The product-space approach also relies on defining so-called pseudo-priors for the
within-model parameters, upon which algorithm efficiency is crucially dependent,
and this can be difficult in practice. Our methods do not involve the need to introduce
such pseudo-priors, although for some missing data problems we need to specify similar
prior distributions for the missing data.

Second, computational methods for the Bayesian analysis of mixture models are well-established,
both when the number of components in the mixture is known \citep{DieRob94}
and when it is not \citep{RichGreen97}. The typical situation under consideration
is one in which the data are assumed to comprise independent and identically distributed
observations, each of which comes from the proposed mixture distribution(s).
In contrast, for our methods we consider only one observation from the mixture model, but this single
observation consists of all the observed data. In other words, rather than assuming each data
point can individually come from any model in the mixture, we assume all data points come
from one of the models in the mixture. This assumption is completely natural when the data are
observations from a stochastic process and do not consist of independent points. However, our
methods apply equally well to independent and identically distributed observations, as illustrated
in the sequel.

The paper is structured as follows. Section 2 contains general theory which describes the
inference framework in detail, and computational matters are described in Section 3. In Section 4
we present a number of examples to illustrate the methods. We conclude with discussion in Section 5.

\section{General Theory}\label{theory}
\subsection{Preliminaries}
In this section we introduce the underlying framework of interest. For ease of exposition we adopt
the usual abuse of notation and terminology in which `a density $\pi(\theta)$' can refer to both the
density function $\pi$ of a random variable $\theta$, or the same function evaluated at a typical
point $\theta$.

\subsection{Mixture model with no missing data}
Suppose we observe data $x$, and wish to consider $n$ competing models $M_1, \ldots, M_n$.
For $i=1, \ldots, n$ denote the probability density of $x$ under model $i$ by $\pi_i( x | \theta_i)$, where
$\theta_i$ denotes the vector of within-model parameters, and set $\theta = (\theta_1, \ldots, \theta_n)$.
We assume that all the $\pi_i( x | \theta_i)$ are densities with respect to the same common reference measure.
Define a mixture model by
\begin{equation}
\label{model}
\pi(x | \alpha, \theta) = \sum_{i=1}^n \alpha_i \pi_i (x | \theta_i),
\end{equation}
where $\alpha = (\alpha_1, \ldots, \alpha_n)$ satisfies $\sum_{i=1}^n \alpha_i = 1$ and $\alpha_i \geq 0$
for $i=1, \ldots, n$.

\subsection{Mixture model with missing data}
In our setting, the data $x$ may be a partial observation of a stochastic process. In consequence,
$\pi_i(x | \theta_i)$  in (\ref{model}) may be intractable, meaning that it cannot be
analytically or numerically evaluated in an efficient manner.
We adopt data augmentation to overcome this problem, as follows.
Let $y = ( y_1, \ldots, y_N )$ be a vector comprising different kinds of `missing data',
and for $i = 1, \ldots, n$ let $\mathcal{I}(i) \subseteq \left\{ 1, \ldots, N \right\}$
and define $y_{\mathcal{I}(i)}$ as the vector with components $y_j$, $j \in \mathcal{I}(i)$.
Thus $y_{\mathcal{I}(i)}$ denotes the missing data for model $i$, in practice usually
chosen to make the augmented probability density $\pi_i(x, y_{\mathcal{I}(i)} | \theta_i)$ tractable.
If model $i$ does not require missing data, then $y_{\mathcal{I}(i)}$ is null.
Note that this formulation allows different models to share common elements of missing data.
Conversely, if each model has its own missing data then we simply set $\mathcal{I}(i) = i$
for $i = 1, \ldots, n$.

In order to define a mixture model using missing data, it is necessary to introduce additional
terms so that each component of the mixture is a probability density function on the possible values of $x$
and $y$. To this end, we assume that there exist tractable probability densities
$\pi_i(y_{- \mathcal{I}(i)} | x, y_{\mathcal{I}(i)}, \theta)$, where $y_{- \mathcal{I}(i)}$
denotes the vector with components $y_j$, $j \notin \mathcal{I}(i)$.
If the latter set is empty then we set $\pi_i(y_{- \mathcal{I}(i)} | x, y_{\mathcal{I}(i)}, \theta) = 1$.
We refer to the $\pi_i(y_{- \mathcal{I}(i)} | x, y_{\mathcal{I}(i)}, \theta)$ terms as {\em missing data prior densities}.
In practice, they need not explicitly depend on any of $x$, $y_{\mathcal{I}(i)}$ or $\theta$, depending on
the application at hand.

Define an augmented mixture model by
\begin{equation}
\label{model_defn}
\pi(x, y | \alpha, \theta) = \sum_{i=1}^n \alpha_i \pi_i (x, y_{\mathcal{I}(i)} | \theta_i) \pi_i(y_{- \mathcal{I}(i)} | x, y_{\mathcal{I}(i)}, \theta).
\end{equation}
Here we assume that each $\pi_i (x, y_{\mathcal{I}(i)} | \theta_i) \pi_i(y_{- \mathcal{I}(i)} | x, y_{\mathcal{I}(i)}, \theta$) term in the sum in (\ref{model_defn}) is a probability
density with respect to a common reference measure, from which it follows that $\pi(x,y | \alpha, \theta)$ is also a probability density.


\subsection{Bayes Factors}
We now show how Bayes factors can be computed directly from certain summaries
of the posterior distribution of $\alpha$ given the data $x$. We assume
that $\alpha$ and $\theta$ are independent {\em a priori}, and that the prior density
$\pi(\theta)$ has marginal densities $\pi_i (\theta_i)$, $i=1, \ldots, n$, which are equal
to the desired within-model prior densities.

By Bayes' Theorem,
\[
\pi ( \alpha | x)  =  \frac{\pi(x|\alpha) \pi(\alpha)}{\pi(x)}
 =  \frac{ \pi(\alpha) \sum_{i=1}^n \alpha_i m_i(x)}{\pi(x)},
\]
where $\pi(\alpha)$ denotes the prior density of $\alpha$ and, for $i=1, \ldots, n$,
\[
m_i(x) = \int \pi_i (x, y_{\mathcal{I}(i)} | \theta_i) \pi(\theta) \, d\theta dy_{\mathcal{I}(i)} =
\int \pi_i (x, y_{\mathcal{I}(i)} | \theta_i) \pi_i(\theta_i) \, d\theta_i dy_{\mathcal{I}(i)},
\]
integrating over all $\theta_j$ with $j \neq i$.
Note also that
\[
1 = \int \pi( \alpha | x )\, d\alpha = \pi(x)^{-1} \sum_{i=1}^n E[\alpha_i] m_i(x),
\]
whence
\begin{equation}
\label{pi(x)}
\pi(x) = \sum_{i=1}^n E[\alpha_i] m_i(x).
\end{equation}

Now for $i \neq j$, the Bayes factor in favour of $M_i$ relative to $M_j$ is defined to be
$B_{ij} = B_{ij}(x) = m_i(x)/m_j(x)$. However,
\begin{eqnarray*}
E [ \alpha_i | x ] & = & \int \alpha_i \pi (\alpha | x) \, d \alpha\\
& = & \pi(x)^{-1} \int \alpha_i \left( \sum_{j=1}^n \alpha_j m_j(x) \right) \pi (\alpha) \, d \alpha \\
& = & \pi(x)^{-1} \sum_{j=1}^n  E[\alpha_i \alpha_j] m_j(x),
\end{eqnarray*}
which combined with (\ref{pi(x)}) yields that
\begin{equation}
\label{E(a|x)}
E[\alpha_i | x] = \frac{ \sum_{j=1}^n  E[\alpha_i \alpha_j] m_j(x)}{ \sum_{j=1}^n E[\alpha_j] m_j(x)}, \; \; \; i=1, \ldots, n.
\end{equation}
Next, fix $k \in \left\{ 1, \ldots, n \right\}$. Dividing the numerator and denominator of the fraction in (\ref{E(a|x)})
by $m_k(x)$ and rearranging we obtain
\begin{equation}
\label{BF_eqn}
\sum_{j=1}^n ( E[\alpha_j] E[\alpha_i | x] - E[\alpha_i \alpha_j]) B_{jk}(x) = 0, \; \; \; i=1, \ldots, n.
\end{equation}

It remains to solve equations (\ref{BF_eqn}) to find $B_{jk}(x)$, $j=1, \ldots, n$. Define $A$ as the
$n \times n$ matrix with elements
\[
A_{ij}  =  E[\alpha_i | x] E [ \alpha_j] - E[\alpha_i \alpha_j], \; \; \; 1 \leq i,j \leq n.\\
\]
Note that $A$ depends on $x$, although we suppress this dependence in our notation.
Define $\tilde{A}_{-k}$ as the $(n-1) \times (n-1)$ matrix formed by removing the
$k$th row and $k$th column of $A$. Similarly for $j \neq k$ define $\tilde{A}_{-jk}$ as the $(n-1) \times (n-1)$ matrix formed
from $\tilde{A}_{-k}$ by replacing the elements $A_{ij}$ with $-A_{ik}$, $i = 1, \ldots, n$, $i \neq k$.\\

{\bf Lemma 1}
(a) If det $\tilde{A}_{-k} \neq 0$ then
\begin{equation}
\label{lemma1}
B_{jk}(x) = \frac{\mbox{det $\tilde{A}_{-jk}$}}{\mbox{det $\tilde{A}_{-k}$}}.
\end{equation}
(b) Suppose that $0 < m_i(x) < \infty$ for $i = 1, \ldots, n$. Then if either (i) $n=2$ and $0 < E[\alpha_1] < 1$,
or (ii) $\alpha$ has a Dirichlet prior distribution, $\mathcal{D}(p_1, \ldots, p_n)$, then
\[
B_{jk}(x) = \frac{A_{jk}}{A_{kj}}.
\]

The proof of Lemma 1 is in the Appendix. The result shows that the required Bayes factors
can be expressed in terms of the prior distribution summaries $E[\alpha_i]$ and $E[\alpha_i \alpha_j]$ and the posterior
means $E[\alpha_i | x]$, $i,j = 1, \ldots, n$.

The condition on the determinant of $\tilde{A}_{-k}$ in Lemma 1(a) is not vacuous in general, as illustrated by the somewhat
pathological case where $\alpha_i$ has a point mass prior for all $i=1, \ldots, n$. Then for all $1 \leq i,j \leq n$,
$E[\alpha_i|x] = E[\alpha_i]$ and $E[\alpha_i \alpha_j]$ = $E[\alpha_i] E[ \alpha_j]$, whence $A_{ij} = 0$ and
(\ref{BF_eqn}) cannot be solved to find the Bayes factors.

At first sight the need for a Dirichlet prior on $\alpha$ to yield simple evaluation of the Bayes factors via Lemma 1 may
appear restrictive. We make three remarks. First, the mixture construction is itself introduced solely as a tool for evaluation
of Bayes factors, and so there is no particular need to assign an arbitrary prior distribution to $\alpha$. Second, in
practice a Dirichlet prior is both straightforward to use and flexible enough for computational purposes as described below.
Third, it may well be that (\ref{lemma1}) holds for arbitrary prior distributions on $\alpha$, subject to mild constraints
which imply that det$\tilde{A}_{-k} \neq 0$, but this does not appear straightforward to prove.

Finally, the simple form for the Bayes factor in Lemma 1 (b) is not true in general; an example for $n=3$ can be
found in the Appendix.

\subsection{Two competing models}
We give special attention to the case $n=2$ since this is of practical importance. Here
we have $\alpha = (\alpha_1, 1 -  \alpha_1)$ and Lemma 1 yields that
\[
B_{12} = \frac{ E[\alpha_1] - E[\alpha_1^2] - E[\alpha_1 | x](1 - E[\alpha_1])}
{E[\alpha_1]E[\alpha_1|x] - E[\alpha_1^2]}.
\]
It follows that
\[
\frac{E[\alpha_1] - E[\alpha_1^2]}{1 - E[\alpha_1]} \leq E[\alpha_1 | x] \leq \frac{E[\alpha_1^2]}{E[\alpha_1]},
\]
with the upper and lower bounds corresponding to Bayes factors entirely in favour of models 1 and 2,
respectively. A practical consequence is that any numerical estimate of $E[\alpha_1|x]$ lying outside these bounds
must be incorrect.

Under the further assumption that $\pi(\alpha)$ is a uniform density, so that $\alpha_1 \sim U(0,1)$, we obtain
\[
B_{12} = \frac{3 E[\alpha_1|x]-1}{2 - 3E[\alpha_1 | x]},
\]
\[
\pi(\alpha_1 | x) \propto \alpha_1 m_1(x) + (1- \alpha_1)m_2(x),
\]
$E[\alpha_1 | x] = (2 m_1(x) + m_2(x))/(3(m_1(x)+m_2(x))$ and $1/3 \leq E[\alpha_1 | x] \leq 2/3$.

Finally, if $\alpha$ is assigned a Dirichlet prior distribution, bounds for $E[\alpha_i | x]$
for any value of $n$ can be obtained. Full details can be found in the proof of Lemma 1 in the Appendix.

\section{Computation}
\subsection{Preliminaries}
We now describe how to use the mixture framework in practice, specifically via Markov chain Monte Carlo methods.
Our objective is to sample from the target density
\begin{equation}
\label{target}
\pi (\alpha, \theta, y | x) \propto \pi(x,y | \alpha, \theta) \pi (\alpha) \pi(\theta),
\end{equation}
and the first issue is that of assigning any missing data prior density terms in $\pi(x,y | \alpha, \theta)$.

\subsection{Missing data prior densities}
Although the desired Bayes factors are invariant to the choice of any missing data prior densities,
this choice is important in practice for computations. This is largely a problem-specific issue, but
we make two general remarks.
First, if all models share the same missing data ($y_1$, say) then no missing data prior
densities are required, and (\ref{target}) becomes
\[
\pi (\alpha, \theta, y | x) \propto \sum_{i=1}^n \alpha_i \pi_i(x,y_1 | \theta_i) \pi (\alpha) \pi(\theta).
\]
Second, it can be beneficial to assign missing data priors which mimic the marginal density of
the $y_{- \mathcal{I}(i)}$ components in other models. As discussed below, the mixing properties of
suitable Markov chain Monte Carlo algorithms are improved if the chains can easily move between different models, and
such movement is hindered if the density of the missing data in one model is very different to
the missing prior density assigned in another.

\subsection{Markov chain Monte Carlo methods}
Sampling from the target density defined at (\ref{target}) will typically be possible via a range of standard
Markov chain Monte Carlo methods, but here we offer some observations on practical aspects.
The fact that the target density is a sum will usually make direct Gibbs sampling infeasible, but the
approach of \cite{DieRob94}, which relies on the introduction of allocation variables which indicate the
`true' model as described in \cite{Demp77}, can be adapted as follows.

Introduce $z = (z_1, \ldots, z_n)$ such that $z_i \in \left\{ 0, 1 \right\}$ and $\sum_{i=1}^n z_i = 1$. Thus
$z$ can take $n$ possible values, each of which is a vector of zeroes other than a 1 at one position. Define the
augmented likelihood
\[
\pi(z,x,y | \alpha, \theta) = \prod_{i=1}^n (\alpha_i  \pi_i (x, y_{\mathcal{I}(i)} | \theta_i) \pi_i(y_{- \mathcal{I}(i)} | x, y_{\mathcal{I}(i)}, \theta))^{z_i},
\]
so that the augmented likelihood at (\ref{model_defn}) is recovered by summing over $z$. If the
prior distribution on $\alpha$ is Dirichlet, $\mathcal{D}(p_1, \ldots, p_n)$, it follows that $\alpha$ has full conditional distribution
\[
\alpha | \cdots \sim \mathcal{D}(p_1+z_1, \ldots, p_n+z_n).
\]
The full conditional distribution of $z$ is multinomial $\mathcal{M}(1; q_1, \ldots, q_n)$, where the probabilities are given by
\[
q_i \propto \alpha_i  \pi_i (x, y_{\mathcal{I}(i)} | \theta_i) \pi_i(y_{- \mathcal{I}(i)} | x, y_{\mathcal{I}(i)}, \theta), \; \; i=1, \ldots, n.
\]
For $j=1, \ldots, n$, $\theta_j$ has full conditional distribution given by
\[
\pi(\theta_j | \cdots) \propto
\pi (\theta) \pi_i (x, y_{\mathcal{I}(i)} | \theta_i) \pi_i(y_{- \mathcal{I}(i)} | x, y_{\mathcal{I}(i)}, \theta)
\]
where $i$ denotes the current model, i.e. $z_i = 1$. Simplification often occurs in practice: for instance,
if $\theta_1, \ldots, \theta_n$ are independent {\em a priori} and there are no missing data prior densities then we obtain
\[
\pi(\theta_j | \cdots) \propto
\left\{
\begin{array}{ll}
\pi_j (\theta_j) & z_j = 0,\\
\pi_j (x, y_{\mathcal{I}(j)} | \theta_j) \pi_j (\theta_j) & z_j = 1.
\end{array}
\right.
\]
Finally, any missing data component $y_j$, $j=1, \ldots, N$, has full conditional distribution given by
\[
\pi(y_j | \cdots) \propto
\left\{
\begin{array}{ll}
\pi_i (x, y_{\mathcal{I}(i)} | \theta_i) & j \in \mathcal{I}(i),\\
\pi_i (y_{- \mathcal{I}(i)} | x, y_{\mathcal{I}(i)}, \theta_i) & j \notin \mathcal{I}(i),
\end{array}
\right.
\]
where $i$ denotes the current model.

The prior distribution for $\alpha$ can often be chosen to improve the mixing of
the Markov chain. In particular this can be achieved by trying to make the
multinomial distribution of $z$ as close to uniform as possible. To illustrate this, consider
the trivial example with two models in which $\pi_1(x)/ \pi_2(x) = m_1(x)/m_2(x) = B_{12} = 50$. The
full conditional distributions for $z_1$ and $\alpha_1$ are, respectively, $Bern(50 \alpha_1/(50 \alpha_1 + (1-\alpha_1)))$
and $Beta(z_1+p_1, 1 - z_1 + p_2)$, where $Bern$ and $Beta$ respectively denote Bernoulli and Beta distributions. Setting
$p_1 = p_2 = 1$ produced wildly different estimates for $B_{12}$ (87.8, 40.9, 547.1) for three algorithm runs of $10^6$ iterations,
while repeating the exercise with $p_1 = 1$ and $p_2 = 50$ yielded estimates 50.3, 50.3 and 50.7.

It is of course not necessary to use allocation variables, and one can equally use any suitable Markov chain Monte Carlo scheme
for the target density. However, the above illustrates the fact that the full conditional distributions of $\theta_i$
and any missing data will be of mixture form, which has implications for the design of efficient algorithms. Note that
this also illustrates that the marginal densities $\pi( \theta_i | x)$ are mixtures, and in particular are
not the same as those obtained from a single-model analysis, which are proportional to
$\pi_i(x | \theta_i) \pi_i(\theta_i)$. The marginal densities can be either be explored via a standard single-model
Markov chain Monte Carlo algorithm, or by using the allocation variables approach and conditioning the output on $z_i=1$ to obtain
within-model posterior density samples for $\theta_i$.

\subsection{Connections with other approaches}
The framework we adopt is related to that described in \cite{CarChib95} and \cite{God01}, in which the target distribution of
interest is defined over a product space of models and their parameters. In order to clarify the
differences in our approach, consider the simplest possible setting in which we have two models defined
by densities $\pi_1(x | \theta_1)$ and $\pi_2(x | \theta_2)$, and independent within-model prior densities $\pi_1(\theta_1)$
and $\pi_2(\theta_2)$. The framework of \cite{CarChib95} and \cite{God01} introduces a model indicator $k \in \left\{ 1,2 \right\}$
to denote the `current' model. The target density of interest is specified by
\[
\pi(k, \theta_1, \theta_2 | x) \propto \pi_k(x|\theta_k) \pi_k(\theta_k | k) \pi(\theta_{3-k} | \theta_k, k) \pi(k),
\]
where it is necessary to specify $\pi(\theta_{3-k} | \theta_k, k)$, i.e. the `prior' for the non-current model
parameter. Assuming $\theta_1$ and $\theta_2$ to be independent of each other and $k$ gives that
$\pi(\theta_{3-k} | \theta_k, k) = \pi_{3-k} (\theta_{3-k})$.

Conversely, our formulation has target density
\[
\pi ( \alpha, \theta_1, \theta_2 | x) \propto  \pi(\alpha) \pi_1(\theta_1) \pi_2(\theta_2) [ \alpha_1 \pi_1(x | \theta_1) + \alpha_2 \pi_2(x|\theta_2)].
\]
If we adopt the allocation-variable approach, the target density becomes
\[
\pi(z, \alpha, \theta_1, \theta_2 | x) \propto \pi(\alpha) \pi_1(\theta_1) \pi_2(\theta_2) [ \alpha_1 \pi_1(x | \theta_1)]^{z_1}  [\alpha_2 \pi_2(x | \theta_1)]^{z_2},
\]
from which we see that it is the existence of the $\alpha$ parameter which distinguishes our formulation from that of \cite{CarChib95} and \cite{God01}.
Since posterior estimation of $\alpha$ is what enables us to estimate Bayes factors, this difference is an important one.

The general formulation in \cite{God01} also allows each model to potentially share parameters with other models. Specifically, the parameters of
model $k$ will be some subset of a set of parameters $\left\{ \theta_1, \ldots, \theta_N \right\}$. This is similar to the way we have dealt with
missing data, although the two set-ups are not technically equivalent, and in particular one cannot simply treat our missing data as model parameters.
The fundamental difference is that missing data may not always require a prior, whereas model parameters always do. For instance, if a density
$\pi(x | \theta)$ is intractable, then our missing data approach uses Bayes' Theorem in the form $\pi(\theta, y | x) \propto \pi(x,y | \theta) \pi (\theta)$
whereas augmenting with extra model parameter $\psi$ gives $\pi(\theta, \psi|x) \propto \pi (x | \theta, \psi) \pi (\theta, \psi)$.

Finally, note that our approach also enables each model to potentially share parameters with other models, since we allow for arbitrary prior dependency
between $\theta_1, \ldots, \theta_n$. Example 2 below illustrates this idea.

\section{Examples}\label{examples}
In this section we illustrate the theory with two examples of classical data sets comprising independent observations,
and three examples featuring population processes or epidemics.

{\bf Example 1: Non-nested regression models for Pines data}
We consider the well-known model choice problem of assigning non-nested linear
regression models to the pines data set in \cite{Wil59}. These data have been analyzed by
several authors \citep[see, for example,][]{HanBrad01, CarChib95, FrielPet08} in order to compare methods
for estimating Bayes factors.  The data describe the maximum compression
strength parallel to the grain $y_i$, the density $x_i$, and the resin-adjusted density $z_i$ for 42
specimens of radiata pine. The two competing models we consider are
\begin{eqnarray*}
 M_1: \quad & y_i = \alpha + \beta(x_i - \bar{x}) + \epsilon_i, \quad & \epsilon_i \sim N(0, \sigma^2); \nonumber \\
 M_2: \quad & y_i = \gamma + \delta(z_i - \bar{z}) + \eta_i, \quad & \eta_i \sim N(0, \tau^2), \nonumber
\end{eqnarray*}
which in vector notation we write as $Y = X(\alpha, \beta)^T + \epsilon$ and $Y = Z(\gamma, \delta)^T + \eta$.
We assigned identical prior distributions to the papers cited above, i.e. $N(\mu_0, v_0) \equiv N((3000,185)^T, \mbox{diag}(10^6,10^4))$
prior distributions for $(\alpha, \beta)^T$ and $(\gamma, \delta)^T$, and $IG(a_0, b_0) \equiv IG(3, (2 \times 300^2)^{-1})$ prior
distributions for $\sigma^2$ and $\tau^2$, where $IG(a,b)$ denotes an inverse gamma distribution with probability
density function
\[
f(x) = \frac{1}{\exp(1/bx) \Gamma(a) b^a x^{a+1}}.
\]
We assigned a Beta$(100, 1)$ prior distribution for $\alpha_1$. Using the allocation variables approach, parameter updates for a Gibbs sampling algorithm
are as follows, where $\theta_1 = (\alpha, \beta, \sigma^2)$ and $\theta_2 = (\gamma, \delta, \tau^2)$:
\begin{eqnarray*}
\alpha_1 | \cdots & \sim & Beta(100+z_1,2-z_1),\\
z_1 | \cdots & \sim & Bern \left(\frac{\alpha_1 \pi_1(x|\theta_1)}{ \alpha_1 \pi_1(x|\theta_1) + (1-\alpha_1)\pi_2(x|\theta_2)} \right),\\
\theta_i | \cdots & \sim &
\left\{
\begin{array}{ll}
(N(\mu_0,v_0), IG(a_0,b_0)) & z_i = 0,\\
(N(\mu_i, v_i), IG(a_i,b_i)) & z_i = 1,
\end{array}
\right.\\
\end{eqnarray*}
where $v_1 = (v_0^{-1} + \sigma^{-2}X^TX)^{-1}$, $v_2 = (v_0^{-1} + \tau^{-2}Z^TZ)^{-1}$, $\mu_1 = v_1(v_0^{-1}\mu_0 + \sigma^{-2}X^TY)$,
$\mu_2 = v_2(v_0^{-1}\mu_0 + \tau^{-2}Z^TY)$, $a_1 = a_2 = (n/2) + a_0$, $b_1 = [b_0 + (1/2)(Y-X(\alpha, \beta)^T)^T(Y-X(\alpha, \beta)^T)]^{-1}$
and $b_2 = [b_0 + (1/2)(Y-Z(\gamma, \delta)^T)^T(Y-Z(\gamma, \delta)^T)]^{-1}$.

We carried out 100 runs
of our method, this being the same as the number of runs used for the methods described in \cite{FrielPet08}.
\cite{FrielPet08} also compared conventional reversible jump methods (with each model {\em a priori} equally likely),
`corrected' Reversible jump methods (model priors chosen to improve mixing) and two power posterior methods (Serial and Population
Markov chain Monte Carlo). Full details can be found in \cite{FrielPet08}, and for convenience we simply quote the results obtained in
Table \ref{tab:pines}, along with our result. The bias is calculated by comparison with the estimate of 4862 obtained by numerical
integration in \cite{GrOHagan98}. It can be seen that our method is certainly competitive.\\

\begin{table}
\centering
\caption{Pines data set: Comparison of Bayes factors from different methods.}{
\label{tab:pines}
\begin{tabular}{lll}
Method & Bias & Standard Error\\
RJMCMC & 67 & 2678\\
RJ corrected & 9 & 124\\
Power posterior (serial MCMC) & 10 & 132\\
Power posterior (population MCMC) & 22 & 154\\
Mixture method & 10 & 39
\end{tabular}}
\end{table}

{\bf Example 2: Three logistic regression models for Pima Indians data}
The well-known Pima Indians data set consists of diabetes incidence
for $n=532$ Pima Indian women along with data on seven covariates, and can be naturally modelled using
logistic regression. In \cite{FrielWyse12}, two specific models are considered. These models ($M_2$ and $M_3$, here)
have four explanatory variables in common (number of pregnancies, plasma glucose concentration, body mass
index and diabetes pedigree function) while $M_2$ has one additional variable (age). The corresponding
Bayes factor $B_{23}$ is calculated using various different methods in \cite{FrielWyse12}, yielding values
in the range 12.83 to 13.96, the latter value coming from a lengthy reversible jump MCMC run which could
plausibly be regarded as the most reliable estimate.

To illustrate our method for three models we also consider an additional model $M_1$, which has the same explanatory variables
as $M_2$ other than diabetes pedigree function. Defining the vector of binary response data $x = (x_1, \ldots, x_n)$,
for $i = 1, 2, 3$ model $i$ has likelihood
\[
\pi_i(x | \theta_i) = \prod_{j=1}^n p_{i,j}^{x_j}(1-p_{i,j})^{1 - x_j}
\]
where $p_{i,j}$ is the probability of incidence for individual $j$ under model $i$, itself
defined in terms of the model $i$ covariates for individual $j$, $z_{ij} = (1, z_{ij}^{(1)}, \ldots, z_{ij}^{(d_i)})$,
and the model $i$ parameter vector $\theta_i = (\theta_{i0}, \theta_{i1}, \ldots, \theta_{id_i})^T$ via the relation
\[
\log \left( \frac{p_{i,j}}{1 - p_{i,j} } \right) = \theta_i^T z_{ij},
\]
where $d_i$ is the number of explanatory variables in model $i$, so that $d_1 = 3$, $d_2 = 4$ and $d_3 = 5$. Thus the parameters
of the three models under consideration are all of different dimension. The elements $\theta_{ij}$ of $\theta_i$ are each
assigned independent Gaussian prior distributions with mean zero and variance 100.

We estimated Bayes factors as follows. First, we assume {\em a priori} that $\theta_{1j} = \theta_{2j} = \theta_{3j}$ for $j=0, \ldots, 3$, so that
in fact we only have six parameters (namely $\theta_{30}, \ldots, \theta_{35}$) in the mixture model formulation. Although this
assumption is not necessary, it was found to help the mixing of the MCMC algorithm. Thus the target density is
\[
\pi (\alpha, \theta_1, \theta_2, \theta_3 | x)  \propto (\alpha_1 \pi_1( x |  \theta_1)  + \alpha_2 \pi_2( x |  \theta_2)
+ \alpha_3 \pi(x | \theta_3)) \pi (\alpha) \pi (\theta_1, \theta_2, \theta_3)
\]
where $\alpha_3 = 1 - \alpha_1 - \alpha_2$, $\theta_1 = (\theta_{30}, \ldots, \theta_{33})^T$, $\theta_2 = (\theta_{30}, \ldots, \theta_{34})^T$,
$\theta_2 = (\theta_{30}, \ldots, \theta_{35})^T$ and $\pi (\theta_1, \theta_2, \theta_3) = \pi ( \theta_{30}, \ldots, \theta_{35})$
is a product of six Gaussian density functions. We set $\alpha \sim \mathcal{D}(1,1,1)$ {\em a priori}.

We implemented our method using an allocation variable approach with Metropolis-Hastings updates for the $\theta_{3j}$ parameters.
From $10^7$ iterations of the resulting Markov chain we obtained estimates $B_{12} = 0.048$ and $B_{23} = 13.94$. The latter
is in line with the values reported in \cite{FrielWyse12}, especially the reversible jump MCMC estimate of 13.96,
while the former is close to an estimate of 0.042 which we obtained using the Laplace approximation method \citep{TiernyKad86} described in \cite{FrielWyse12}.\\

{\bf Example 3: Poisson process vs. linear birth process}
Our first population process example is analytically tractable and illustrates that our methods produce results
in agreement with the known true values. Consider data given by the vector of event
times $x = (x_1, \ldots, x_n)$ observed during a time interval $[0,T]$, where
$0 \leq x_1 \leq x_2 \leq \ldots \leq x_n \leq T$.  We will compare
two models, namely a homogeneous Poisson process of rate $\lambda$ ($M_1$) and a linear birth process $\left\{ X(t): t \in [0,T]
\right\}$ with per-capita birth rate $\mu$ and $X(0)=1$ ($M_2$). Suppose further that $\lambda$ and $\mu$ are assigned independent exponential prior distributions with mean $\theta^{-1}$. The model likelihoods, which we write as densities with
respect to the reference measure induced by a unit rate Poisson process on $[0,T]$, are
\[
\pi_1(x | \lambda) = \lambda^n \exp\left\{- (\lambda-1) T\right\}, \; \; \pi_2(x | \mu) = n! \mu^n \exp \left\{ -\mu [(n+1)T - S(x)] + T \right\},
\]
where $S(x) = \sum_{j=1}^n x_j$. In this setting no missing data are required so we use the model defined at (\ref{model}).
The Bayes factor in favour of $M_1$ relative to $M_2$ is
\begin{eqnarray*}
B_{12} = \frac{\int \pi_1(x | \lambda) \pi(\lambda) \;d\lambda}{\int \pi_2(x | \mu) \pi(\mu) \;d\mu}
& = & \frac{ \int_0^\infty \theta \lambda^n \exp\left\{- \lambda (T+ \theta) \right\} \; d \lambda}
{ \int_0^\infty \theta n! \mu^n \exp \left\{ -\mu [(n+1)T - S(x) + \theta] \right\} \; d \mu}\\
& = & \frac{ [ (n+1)T - S(x) + \theta]^{n+1}}{(T+\theta)^{n+1} n!}.
\end{eqnarray*}
Assuming that $\alpha_1 \sim U(0,1)$ {\em a priori}, a simple Gibbs sampler for the target density consists
of parameter updates as follows:
\begin{eqnarray*}
\alpha_1 | \cdots & \sim & Beta(z_1+1,2-z_1),\\
z_1 | \cdots & \sim & Bern \left(\frac{\alpha_1 \pi_1(x|\lambda)}{ \alpha_1 \pi_1(x|\lambda) + (1-\alpha_1)\pi_2(x|\mu)} \right),\\
\lambda | \cdots & \sim &
\left\{
\begin{array}{ll}
\Gamma(1, \theta) & z_1 = 0,\\
\Gamma(n+1,T+\theta) & z_1 = 1,
\end{array}
\right.\\
\mu | \cdots & \sim &
\left\{
\begin{array}{ll}
\Gamma(n+1,(n+1)T - S(x) +\theta) & z_1 = 0,\\
\Gamma(1, \theta) & z_1 = 1,
\end{array}
\right.
\end{eqnarray*}
where $\Gamma(m, \xi)$ denotes a Gamma distribution with density $f(x) \propto x^{m-1} \exp(- \xi x)$.

Typical results from algorithm runs are given in Table \ref{tab:ex1}, illustrating that the Gibbs sampler recovers the true known values. We found that the algorithm mixing was good in all cases.

\begin{table}
\centering
\caption{Example 3: Bayes factors from algorithm output ($\hat{B}_{12}$) compared to true values ($B_{12}$).}{
\label{tab:ex1}
\begin{tabular}{cccccc}
$n$ & $T$ & $S(x)$ & $\theta$ & $\hat{B}_{12}$ &  $B_{12}$\\
5 & 10 & 36 & 1 & 1.15 & 1.148\\
5 & 10 & 36 & 0.01 & 1.58 & 1.587\\
5 & 10 & 25 & 1 & 10.25 & 10.239\\
10 & 20 & 150 & 1 & 0.18 & 0.181
\end{tabular}}
\end{table}

Finally, we comment on the relationship between the above algorithm and standard reversible jump methods. The
latter requires a way of proposing a value of $\mu$ given $\lambda$ for jumps from $M_1$ to $M_2$, and vice versa. In
practice it is not immediately obvious how best to do this, but an approach suggested in \cite{Gr03} is to propose
$\mu$ independently of $\lambda$, ideally according to the within-model density $\pi(\mu | x)$. This is similar to what
we obtain above.\\

{\bf Example 4: Epidemic model with two different infection periods}
Recall the standard Susceptible-Infective-Removed epidemic model (see e.g. \cite{AndBritt00}, Chapter 2), defined
as follows. A closed population contains $N+a$ individuals of whom $N$ are initially susceptible and $a$ initially infective.
Each infective remains so for a period of time distributed according to a specified random variable $T_I$, known as the
infectious period, after which it becomes removed and plays no further part in the epidemic. During its infectious period an infective makes contact with each other member of the population at times given
by a homogeneous Poisson process of rate $\beta/N$, and any contact occurring with a susceptible individual results in that
individual immediately becoming infective. The infectious periods of different individuals and the Poisson processes between
different pairs of individuals are assumed to be mutually independent. The epidemic ends when there are no infectives
left in the population.

A distinguishing characteristic of infectious disease data is that the infection process itself is rarely observed,
and so we suppose that the data $r$ consist of $n$ observed removal times $r_1 \leq \ldots \leq r_n$. We consider two
competing models, namely that $T_I \sim \Gamma(1, \gamma)$ ($M_1$) and $T_I \sim \Gamma(m, \lambda)$ ($M_2$), where
the shape parameter $m$ will be assumed known. Both model likelihoods $\pi_1(r | \gamma)$ and $\pi_2(r | \lambda)$ are intractable
in practice since their evaluation relies on integrating over all possible realisations of the infection process,
and so we introduce missing data as follows.

For $j=1, \ldots, n$ define $i_j$ as the infection time of the
individual removed at time $r_j$. We assume that there is $a=1$ initial infective, denoted by $p$, so that
$i_p \leq i_j$ for all $j \neq p$. For simplicity we assume {\em a priori} that $p$ is equally likely to be
any of the $n$ infected individuals and that $i_p$ has an improper uniform prior density on $(- \infty, r_1)$.
Finally, define $i = \left\{ i_j : i \neq p \right\}$ to be the $n-1$
non-initial infection times. For $k=1, 2$ the augmented model likelihoods, which we write here with respect to
Lebesgue measure on $\mathbb{R}^{2n-1}$, are
\begin{eqnarray*}
\lefteqn{\pi_k( i, r | p, i_p, \beta_k, \eta_k) =}\\
&& \left( \prod_{j=1; j \neq p}^{n} (\beta_k/N) I(i_j-) \right)
\exp \left\{ - (\beta_k/N) \int_{i_p}^{r_n} S(t) I(t) \; dt \right\}  \left( \prod_{j=1}^n f_k(r_j-i_j| \eta_k) \right),
\end{eqnarray*}
where $S(t)$ and $I(t)$ denote respectively the numbers of susceptibles and infectives at time $t$, $I(t-) = \lim_{s \uparrow t}I(s)$, $\beta_k$ denotes the parameter $\beta$ under $M_k$, $f_k$ denotes the infectious period density under $M_k$, $\eta_1 = \gamma$ and $\eta_2 = \lambda$ (see e.g. \cite{OnRob99}, \cite{StrefGib04}, \cite{HohleON05}). Note that in this formulation, the missing data $i$, $p$ and $i_p$ are assumed
common to both models, although these quantities could also be model-specific.

The target density of interest is
\[
\pi(\alpha_1, \beta_1, \beta_2, \gamma, \lambda | r) \propto [ \alpha_1 \pi_1( i, r | p, i_p, \beta_1, \gamma)
+ (1-\alpha_1) \pi_2( i, r | p, i_p, \beta_2, \lambda)] \pi(\beta_1) \pi(\beta_2) \pi(\gamma) \pi(\lambda).
\]
Note that here we need no missing data priors densities because the the missing data appear in both model likelihoods.
Prior distributions for $\beta_1, \beta_2, \gamma$ and $\lambda$ were all set as $\Gamma(1, 1)$, and
$\alpha_1 \sim Beta(p_1,p_2)$.

Introducing the allocation variable $z_1$ yields the full conditional distributions below, each of which
yields a simple Gibbs update for the parameter in question.
\begin{eqnarray*}
\alpha_1 | \cdots & \sim & Beta(z_1+p_1,1-z_1 + p_2),\\
z_1 | \cdots & \sim & Bern \left(\frac{\alpha_1 \pi_1( i, r | p, i_p, \beta_1, \gamma)}
{ \alpha_1 \pi_1( i, r | p, i_p, \beta_1, \gamma) + (1-\alpha_1)\pi_2( i, r | p, i_p, \beta_2, \lambda) } \right),\\
\beta_1 | \cdots & \sim &
\left\{
\begin{array}{ll}
\Gamma(n, N^{-1} \int_{i_p}^{r_n} S(t) I(t) \; dt + 1) & z_1 = 0,\\
\Gamma(1, 10^{-3}) & z_1 = 1,
\end{array}
\right.\\
\beta_2 | \cdots & \sim &
\left\{
\begin{array}{ll}
\Gamma(1, 1) & z_1 = 0,\\
\Gamma(n, N^{-1} \int_{i_p}^{r_n} S(t) I(t) \; dt + 1) & z_1 = 1,
\end{array}
\right.\\
\gamma | \cdots & \sim &
\left\{
\begin{array}{ll}
\Gamma(n+1,  \sum_{j=1}^n (r_j - i_j) + 1) & z_1 = 0,\\
\Gamma(1, 1) & z_1 = 1,
\end{array}
\right.\\
\lambda | \cdots & \sim &
\left\{
\begin{array}{ll}
\Gamma(1, 1) & z_1 = 0,\\
\Gamma(nm+1, \sum_{j=1}^n (r_j - i_j) + 1)  & z_1 = 1,
\end{array}
\right.\\
\end{eqnarray*}
Finally, the infection time parameters $i$, $i_p$ and $p$ are updated using a Metropolis-Hastings step
as follows. One of the $n$ infected individuals, $j$ say, is chosen uniformly at random. A proposed
new infection time for $j$ is defined as $i_j^*= r_j - x$, where $x$ is sampled from a $\Gamma(1, \delta)$
distribution. Note that this may also result in proposed new values for $p$ and $i_p$; either way, proposed
values are denoted $i^*$, $i^*_{p^*}$ and $p^*$ and accepted with probability
\[
1 \wedge \frac{\pi_k( i^*, r | p^*, i^*_{p^*}, \beta_k, \eta_k)}{\pi_k( i, r | p, i_p, \beta_k, \eta_k)} \exp(\delta(i_j - i^*_j)),
\]
where $k=2-z_1$ denotes the current model.

To illustrate the algorithm, we considered the Susceptible-Infective-Removed model with $N=50$, $a=1$, various $\beta$ values
and $\Gamma(\tilde{m},\tilde{\lambda})$ infectious periods with three different choices for $(\tilde{m},\tilde{\lambda})$. For
each scenario we simulated 100 data sets, and evaluated the Bayes factor using the above algorithm
for each data set.
For two of the $(\tilde{m},\tilde{\lambda})$ pairs we set the shape parameter $m$ in $M_2$ equal
to $\tilde{m}$, and for one we did not.
In practice, one is rarely interested in data from epidemics with few cases, so we also
evaluated the Bayes factors using a subset of each of the 100 simulations in which the epidemic
had clearly `taken off', evaluated by eye, which we refer to as major epidemics. The numbers of
major epidemics were 63, 56 and 65 for scenarios A, B and C, respectively.

Table \ref{tab:ex2} contains a summary of the Bayes factors estimated from the simulated data sets.
In scenarios A, B and C the true models are
$M_2$, both $M_1$ and $M_2$, and $M_1$ respectively. The estimated Bayes factors behave as we might expect,
giving clear evidence in favour of models $M_2$ and $M_1$ for scenarios A and C respectively, whilst for
scenario B the mean of $B_{12}$ is close to the true value of 1. In scenario A there is a marked
difference in the Bayes factors when using all simulations compared to using only major epidemics.
A possible explanation is that major epidemics contain more data, and so any difference between the models
becomes easier to detect. There is a less pronounced difference in Bayes factors in scenario C, although
the large posterior standard deviations suggest there is no compelling evidence for a clear difference in
this case.\\

\begin{table}
\centering
\caption{Example 4: Bayes factors from algorithm output.}{
\label{tab:ex2}
\begin{tabular}{cccccc}
Scenario & True model & $\beta$ & $M_2$ & $E[B_{12}] (st. dev.)$ & $E[B_{12}] (st. dev.)$\\
&&&& (all simulations) & (major epidemics)\\
A & $\Gamma(5,5)$ & 2 & $\Gamma(5,\lambda)$ & 0.06 (0.06) & 0.008 (0.006)\\
B & $\Gamma(1,0.75)$ & 1 & $\Gamma(1, \lambda)$ &1.03 (0.17) & 1.05 (0.22)\\
C & $\Gamma(1,1)$ & 3 & $\Gamma(2, \lambda)$ & 3022 (3969) & 2291 (3428)\\
\end{tabular}}
\end{table}

{\bf Example 5: Epidemic model vs. Poisson process}
Our final example is motivated by the situation in which we wish to decide whether observed
cases of disease are the result of an epidemic (with transmission between individuals) or
simply sporadic events. Specifically, suppose we observe $n$ events at times $0 < r_1 < \ldots < r_n < T$,
and let $r = (r_1, \ldots, r_n)$. Under model $M_1$, $r$ is a vector of event times of a homogeneous Poisson process
of rate $\lambda$ observed during the time interval $[0,T]$. Under model $M_2$, $r$ is a vector of
removal times in a Susceptible-Infective-Removed epidemic model with exponentially distributed infectious periods, again observed during $[0,T]$.

As for the previous example, we proceed by adding unobserved infection times in order to obtain a
tractable likelihood for $M_2$. For simplicity we assume that there is one initial infective at time zero,
and furthermore that there is a population of $N$ individuals in total, where $N \geq n$. Unlike the
previous example, in which we unobserved infection times with observed removal times, we here
define $i = (i_2, \ldots, i_m)$ to be a vector of $m$ ordered infection times, so
that $0 = i_1 < i_2 < \ldots, < i_m$, where $n \leq m \leq N$. The reason for this approach is that it
appears to be easier when it comes to assigning missing data prior densities, as described below.
Note also that under $M_2$ we allow the possibility that the epidemic is still in progress at time $T$.

The likelihood for $M_1$ and augmented likelihood for $M_2$ are respectively given by
\begin{eqnarray*}
\pi_1(r | \lambda) &  = & \lambda^n \exp\left\{- \lambda T\right\}, \\
\pi_2(i,r | \beta, \gamma) & = &
\left( \prod_{j=2}^{m} \beta S(i_j-) I(i_j-) \right)
\left( \prod_{j=1}^n  \gamma I(r_j-) \right)
\exp \left\{ - \int_{0}^{T} \beta S(t) I(t) + \gamma I(t) \; dt \right\}.
\end{eqnarray*}
To proceed we require a missing data prior density $\pi_1(i | r, \lambda, \beta, \gamma)$. Now
for a given ordered vector of event times $r$, $\pi_2(i,r | \beta, \gamma) > 0$ if and only if
$i \in \mathcal{F}(r)$, where
\[
\mathcal{F}(r) = \left\{ i : i_1 < i_2 < \ldots < i_m < T; i_k < r_{k-1}, k = 1, \ldots, n+1; i_k < T, k=n+2, \ldots, m \right\}.
\]
One way to define $\pi_1(i | r, \lambda, \beta, \gamma)$ is via the following construction, which
simulates an element of $\mathcal{F}(r)$.
First, select $m$ according to some probability mass function $f$ on $\left\{ n, n+1, \ldots, N \right\}$.
Next, sequentially set $i_2 \sim TrExp(\mu; i_1, r_1)$, $i_3 \sim TrExp(\mu; i_2, r_2), \ldots$,
$i_{n+1} \sim  TrExp(\mu; i_n, r_n)$, $i_{n+2} \sim TrExp(\mu; i_{n+1}, T), \ldots$, $i_m \sim
TrExp(\mu; i_{m-1}, T)$, where $TrExp(\mu; a,b)$ denotes an exponential random variable with rate $\mu$,
truncated to the interval $(a,b)$.
This in turn induces a probability
distribution with density
\[
\pi_1(i | r) = f(m) \prod_{j=1}^{m-1} \frac{\mu \exp(- \mu i_j)}{\exp(- \mu i_{j-1}) - \exp(- \mu s_{j-1}) }, \; \; \; \; \; i \in \mathcal{F}(r),
\]
where $s_j = r_j$ for $j=1, \ldots, n$ and $s_j = T$ for $n < j \leq m$, and we set
$\pi_1(i | r, \lambda, \beta, \gamma) = \pi_1(i | r)$.

We remark that it is not necessary to define the missing data prior density in this manner. For instance,
one could proceed by choosing $m$ as before and then assigning a uniform density to the set
$\left\{ i : i_2 < i_3 < \ldots < i_m \right\}$. The practical drawback with this is that, if using
allocation variables, the Markov chain can never leave model $M_1$ if $i$ is such that
$\pi_2(i,r | \beta, \gamma) = 0$.

Prior distributions were assigned as $\beta \sim \Gamma(\nu_\beta,\mu_\beta)$, $\gamma \sim \Gamma(\nu_\gamma, \mu_\gamma)$,
$\lambda \sim \Gamma(\nu_\lambda, \mu_\lambda)$ and $\alpha_1 \sim Beta(p_1,p_2)$.

A Markov chain Monte Carlo algorithm is easily developed in a similar manner to the previous example. Specifically we
have the following full conditional distributions:
\begin{eqnarray*}
\alpha_1 | \cdots & \sim & Beta(z_1 + p_1,1-z_1+p_2),\\
z_1 | \cdots & \sim & Bern \left(\frac{\alpha_1 \pi_1(r | \lambda) \pi_1(i|r)}
{ \alpha_1 \pi_1(r | \lambda) \pi_1(i|r) + (1-\alpha_1)\pi_2( i, r | \beta, \gamma) } \right),\\
\lambda | \cdots & \sim &
\left\{
\begin{array}{ll}
\Gamma(\nu_\lambda, \mu_\lambda) & z_1 = 0,\\
\Gamma(n+\nu_\lambda, T + \mu_\lambda) & z_1 = 1,
\end{array}
\right.\\
\beta | \cdots & \sim &
\left\{
\begin{array}{ll}
\Gamma(m - 1 + \nu_\beta, \int_{0}^{T} S(t) I(t) \; dt + \mu_\beta) & z_1 = 0,\\
\Gamma(\nu_\beta, \mu_\beta) & z_1 = 1,
\end{array}
\right.\\
\gamma | \cdots & \sim &
\left\{
\begin{array}{ll}
\Gamma(n+\nu_\gamma, \int_{0}^{T} I(t) \; dt + \mu_\gamma) & z_1 = 0,\\
\Gamma(\nu_\gamma, \mu_\gamma) & z_1 = 1.
\end{array}
\right.\\
\end{eqnarray*}
Updates for $i$ are achieved as follows. If $z_1=1$ then $i$ has full conditional density
$\pi_1(i | r)$ which can be sampled as described above. If $z_1=0$ then $i$ can be updated
by moving, adding or deleting infection times as described in \cite{OnRob99}.

To illustrate this algorithm we considered a data set taken from an outbreak of Gastroenteritis described
in \cite{BrittONeill02} which take the form of 28 case detection times among a population of 89 individuals.
The daily numbers of cases on days 0 to 7 are given respectively by
\[
1,0,4,2,3,3,10,5.
\]
Strictly speaking, such data should be analysed by allowing the unknown time of the initial infection, $i_1$, to
be estimated (see e.g. \cite{OnRob99}). Since our main objective here is to illustrate our methodology, we instead
make the simplifying assumption that day 0 actually corresponded to the start of the outbreak, and then consider
the remaining 27 case detection times.

For the missing data prior density $\pi_1(i | r)$ we set
\[
f(m) = \frac{(1 - \theta)^{m-n} \theta} { 1 - (1-\theta)^{N-n+1}}, \; \; \; m = n, \ldots, N,
\]
so that $m$ has a truncated Geometric distribution with parameter $\theta$, and set $\mu = 4$ in the truncated
exponential distribution.

We ran the algorithm with two choices of $T$, the time of observation, with $\beta$, $\gamma$ and $\mu$ all given
$Exp(1)$ prior distributions. First, with $T=10$ we estimated the Bayes factor in
favour of the Poisson model to be 0.003, here using $p_1 = 400, p_2 = 1$ to obtain reasonable mixing in the algorithm.
So in this case there appears to be overwhelming evidence to suggest that the case detection times are better described by
an epidemic model than a Poisson process. Second, we set $T=3.5$ and used only the case observation times up until day 3.
We estimated the Bayes factor in favour of the Poisson model to be 21.1. In comparison to the $T=10$ case we would certainly
expect a value closer to 1, since there are less data, and equally it is intuitively reasonable that there are insufficient
data to provide evidence in favour of an epidemic.

\section{Discussion}
We have presented a new method for evaluating Bayes factors. Although motivated by epidemic
models and population processes, our approach is clearly applicable in more general settings,
as illustrated by the pines data set and Pima Indians data set examples in Section 4. The methods
permit data imputation as necessary, and can cater for models which share common parameters.

The methods we propose are not without drawbacks. First,
in common with the product-space methods it seems likely that they are best suited to situations in which there are only a small
number of competing models, although we have not investigated this issue in this paper. Second, constructing missing data prior
densities, when required, seems likely to require problem-specific insights in order to obtain reasonably efficient algorithms.
Intuitively we expect that it is best to choose missing data prior distributions to mimic the true distribution of missing
data in competing models. These aspects, as well as the method in general, appear worthy of more detailed exploration.\\[2ex]

\section*{Acknowledgement}
We thank Christian Robert and Andy Wood for helpful discussions about this work.
We acknowledge funding from UK Engineering and Physical Sciences Research Council.

\appendix
\section*{Appendix}

{\bf Proof of Lemma 1}
(a) Define
\[
\mathbf{b} = [ B_{1k}(x) \cdots B_{nk}(x)]^T,
\]
so that (\ref{BF_eqn}) can be written as the matrix equation $A\mathbf{b} = \mathbf{0}$.
Since $B_{kk}(x) = 1$, we can rewrite (\ref{BF_eqn}) as
\begin{equation} \label{BF_eqn2}
\sum_{j \neq k} A_{ij} B_{jk}(x) = - A_{ik}, \; \; \; i=1, \ldots, n.
\end{equation}
Now, for $1 \leq l,j \leq n$,
\begin{eqnarray}
A_{lj} & = & E[\alpha_l | x] E[\alpha_j] - E[\alpha_l \alpha_j] \nonumber \\
& = & E[\alpha_j] \left( 1 - \sum_{i \neq l} E[\alpha_i | x] \right) - E \left[ \left( 1 - \sum_{i \neq l} \alpha_i \right) \alpha_j \right] \nonumber \\
& = & - \sum_{i \neq l} \left( E[\alpha_j] E[\alpha_i | x] - E[\alpha_i \alpha_j] \right) \nonumber \\
& = & - \sum_{i \neq l} A_{ij}.
\label{BF_eqn3}
\end{eqnarray}
Summing (\ref{BF_eqn2}) over $i \neq k$ and using (\ref{BF_eqn3}) now yields
\begin{eqnarray*}
\sum_{i \neq k} \sum_{j \neq k} A_{ij} B_{jk}(x) & = & - \sum_{i \neq k} A_{ik}\\
\mbox{so} \; \; \; \sum_{j \neq k} \left( \sum_{i \neq k} A_{ij} \right) B_{jk}(x) & = & - \sum_{i \neq k} A_{ik}\\
\mbox{so} \; \; \; \sum_{j \neq k} A_{kj} B_{jk}(x) & = & - A_{kk},
\end{eqnarray*}
which is the equation obtained from (\ref{BF_eqn2}) when $i=k$. In other words, at least the $k$th equation in
(\ref{BF_eqn2}) is redundant. It is therefore sufficient to consider the system of equations defined by
\begin{equation} \label{Ab=c}
\tilde{A}_{-k} \tilde{\mathbf{b}} = \tilde{\mathbf{c}},
\end{equation}
where $\tilde{\mathbf{b}}$ is
the $(n-1) \times 1$ column vector formed by removing $B_{kk}(x)=1$ from $\mathbf{b}$, and $\tilde{\mathbf{c}}$
is the $(n-1) \times 1$ column vector with components $-A_{ik}$ for $i = 1, \ldots, n$, $i \neq k$. Application of
Cramer's rule to solve (\ref{Ab=c}) now yields part (a).

(b) For the second part, we require some preliminary results. Write $E[\alpha_i | x] = f(m_i(x))$, say. From (\ref{E(a|x)}) we have
\begin{eqnarray*}
f(m_i(x)) & = & \frac{ \sum_{j=1}^n  E[\alpha_i \alpha_j] m_j(x)}{ \sum_{j=1}^n E[\alpha_j] m_j(x)}\\
& = & \frac{ \sum_{j \neq i}  E[\alpha_i \alpha_j] m_j(x) + E[\alpha_i^2] m_i(x)}{\sum_{j \neq i} E[\alpha_j] m_j(x)  + E[\alpha_i]m_i(x)}.
\end{eqnarray*}
Differentiation yields that $f'(m_i(x)) \geq 0$ if and only if $C \geq 0$, where
\[
C = \sum_{j \neq i} (E[\alpha_i^2]E[\alpha_j] - E[\alpha_i]E[\alpha_i \alpha_j]) m_j(x).
\]
Thus if $C \geq 0$ we obtain the bounds
\begin{equation}
\label{AppBounds}
\frac{\sum_{j \neq i} E[\alpha_i \alpha_j] m_j(x)}{\sum_{j \neq i} E[\alpha_j ] m_j(x)} \leq E[\alpha_i | x] \leq \frac{E[\alpha_i^2]}{E[\alpha_i]},
\end{equation}
and moreover the lower and upper bounds are attained when $m_i(x) = 0$ and $m_i(x) \rightarrow \infty$, respectively. In particular,
for $C > 0$ and $0 < m_i(x) < \infty$ then both inequalities are strict. If $C \leq 0$ then the inequalities in (\ref{AppBounds}) are simply reversed.
From now on we assume that $0 < m_i(x) < \infty$ for all $i = 1, \ldots, n$.

Now if $n=2$ then (\ref{AppBounds}) yields that for $i \neq j$,
\[
\frac{E[\alpha_1 \alpha_2]}{E[\alpha_j ]} \neq E[\alpha_i | x],
\]
from which it follows that $A_{ij} = E [ \alpha_i | x ] E [ \alpha_j ] - E [ \alpha_1 \alpha_2] \neq 0$. The result for $n=2$ now follows
directly from part (a).

For the final part, in which $\alpha$ has a Dirichlet prior distribution, we first show that det $\tilde{A}_{-k} \neq 0$, so
that (\ref{Ab=c}) has a unique solution. Secondly we show that this solution is given by $B_{jk}(x) = A_{jk} / A_{kj}$.

We start with conditions under which $C>0$. Specifically, if Cov$(\alpha_i,\alpha_j)<0$ for all $i \neq j$ and Var$(\alpha_i)>0$ then
\begin{eqnarray*}
E[\alpha_i]E[\alpha_j] & > & E[\alpha_i \alpha_j]\\
\mbox{so} \; \; \; E[\alpha_i^2] E[\alpha_i]E[\alpha_j] & > & E[\alpha_i]^2 E[\alpha_i \alpha_j],
\end{eqnarray*}
from which it follows that $C > 0$.

Next, suppose that $\alpha \sim \mathcal{D}(p_1, \ldots, p_n)$ and set $p_0 = \sum_{i=1}^n p_i$. Thus for $i \neq j$,
$E[\alpha_i \alpha_j] = p_ip_j/(p_0(p_0+1))$, $E[\alpha_i] = p_i/p_0$, $E[\alpha_i^2] = p_i(p_i+1)/p_0(p_0+1)$, Cov$(\alpha_i,\alpha_j)<0$ and Var$(\alpha_i)>0$.
It follows that $C > 0$ and that (\ref{AppBounds}) simplifies to
\begin{equation}
\label{DirBounds}
\frac{p_i}{p_0+1} < E[\alpha_i | x] < \frac{p_i+1}{p_0+1}.
\end{equation}

Next, note that for $i \neq j$ we have
\begin{eqnarray*}
A_{ij} & = & E[\alpha_j]E[\alpha_i | x] - E[\alpha_i \alpha_j]\\
& = & \frac{p_j}{p_0} \left( E[\alpha_i | x] - \frac{p_i}{p_0+1} \right)\\
& = & b_j a_i(x),
\end{eqnarray*}
say, where $b_j = p_j/p_0$. It follows from (\ref{DirBounds}) that $A_{ij} > 0$.
Similarly
\begin{eqnarray*}
A_{ii} & = &  \frac{p_i}{p_0} \left( E[\alpha_i | x] - \frac{p_i+1}{p_0+1} \right)\\
& = & b_i \tilde{a}_i(x),
\end{eqnarray*}
say. Recall that $\tilde{A}_{-k}$ is the matrix $A$ with the $k$th row and column deleted. It now
follows that
\[
\mbox{det}( \tilde{A}_{-k} ) = \left( \prod_{i \neq k} b_i \right) \mbox{det}(D + E),
\]
where $D$ is an $(n-1) \times (n-1)$ diagonal matrix with entries $\tilde{a}_i(x) - a_i(x) = -1/(p_0+1)$, $i \neq k$, and $E$ is
an $(n-1) \times (n-1)$ matrix consisting of $(n-1)$ identical columns, each of which contains the $(n-1)$ entries
$a_i$, $i \neq k$. Moreover we can write $E$ as the product $u v^T$, where $u$ is an $(n-1) \times 1$ column vector
with entries $a_i$, $i \neq k$, and $v$ is the $(n-1) \times 1$ column vector of 1's. In particular, det $\tilde{A}_{-k} \neq 0$
if and only if det$(D+E)  = \mbox{det}(D + uv^T) \neq 0$.

Now from the matrix determinant lemma,
\[
\mbox{det}(D + uv^T) = (1 + v^T D^{-1}u)\mbox{det}(D),\
\]
and since det$(D) = (-1/(p_0+1))^{n-1} \neq 0$, we focus on $1 + v^T D^{-1}u$.
Now
\begin{eqnarray*}
1 + v^T D^{-1}u & = & 1 - \sum_{i \neq k} (p_0+1) a_i(x)\\
& = & 1 - \sum_{i \neq k} [ (p_0+1)E[\alpha_i |x] - p_i]\\
& = & 1 - (p_0+1)(1 - E[\alpha_k | x]) + (p_0-p_k)\\
& = & E[\alpha_k | x](p_0+1) - p_k >0,
\end{eqnarray*}
where the last inequality follows from (\ref{DirBounds}). Hence det $\tilde{A}_{-k} \neq 0$ as required.

Finally, we show that for $i \neq k$, (\ref{BF_eqn2}) is satisfied by $B_{jk}(x) = A_{jk} / A_{kj}$. First, it is
straightforward to show that for $i \neq k$,
\begin{equation}
\label{a_sum}
\tilde{a}_i(x) + \sum_{j \neq k; j \neq i} a_j(x) = - a_k(x).
\end{equation}
Now,
\begin{eqnarray*}
\sum_{j \neq k} A_{ij} \frac{A_{jk}}{A_{kj}} & = & b_i \tilde{a}_i(x)
\frac{b_k a_i(x)}{b_i a_k(x)} + \sum_{j \neq k; j \neq i} b_j a_i(x) \frac{b_k a_j(x)}{b_j a_k(x)}\\
& = & \frac{a_i(x) b_k}{a_k(x)} \left( \tilde{a}_i(x) + \sum_{j \neq k; j \neq i} a_j(x) \right)\\
& = & \frac{a_i(x) b_k}{a_k(x)} ( -a_k(x))\\
& = & -A_{ik},
\end{eqnarray*}
using (\ref{a_sum}). Hence for $i \neq k$, (\ref{BF_eqn2}) is satisfied by $B_{jk}(x) = A_{jk} / A_{kj}$ as required.\\

\mbox{}\; \;{\em Example} A1
To illustrate the calculations in part (a), consider the case $n=3$, $k=1$. The equation $A\mathbf{b} = \mathbf{0}$ is
\[
\left[
\begin{array}{ccc}
A_{11} & A_{12} & A_{13}\\
A_{21} & A_{22} & A_{23}\\
A_{31} & A_{32} & A_{33}
\end{array}
\right]
\left[
\begin{array}{c}
1\\
B_{21}\\
B_{31}
\end{array}
\right]
=
\left[
\begin{array}{c}
0\\
0\\
0
\end{array}
\right] ,
\]
and so
\[
\tilde{A}_{-1} =
\left[
\begin{array}{cc}
A_{22} & A_{23}\\
A_{32} & A_{33}
\end{array}
\right] , \; \;
\tilde{\mathbf{b}} =
\left[
\begin{array}{c}
B_{21}\\
B_{31}
\end{array}
\right] , \; \;
\mathbf{c} =
\left[
\begin{array}{c}
-A_{21}\\
-A_{31}
\end{array}
\right] .
\]
Applying Lemma 1 yields
\[
B_{21} = \frac{\mbox{det $\tilde{A}_{-21}$}}{\mbox{det $\tilde{A}_{-1}$}} =
\frac{\mbox{det $
\left[
\begin{array}{cc}
-A_{21} & A_{23}\\
-A_{31} & A_{33}
\end{array}
\right]
$}}{\mbox{det $\tilde{A}_{-1}$}}, \; \; \; \;
B_{31} = \frac{\mbox{det $\tilde{A}_{-31}$}}{\mbox{det $\tilde{A}_{-1}$}} =
\frac{\mbox{det $
\left[
\begin{array}{cc}
A_{22} & -A_{21}\\
A_{32} & -A_{31}
\end{array}
\right]
$}}{\mbox{det $\tilde{A}_{-1}$}}.
\]

\mbox{}\;\;{\em Example} A2
To show that $B_{jk}(x)$ does not equal $A_{jk}/A_{kj}$ in general, suppose that $n=3$, that $m_i(x) = i$ for $i=1,2,3$,
and that $\alpha$ has a mixed Dirichlet prior distribution given by
\[
\alpha \sim (0.5) \mathcal{D}(1,1,1) + (0.5) \mathcal{D}(1,2,1).
\]
Direct calculation then yields that $E [ \alpha_1 ] = E [ \alpha_3] = 7/24$, $E [ \alpha_2 ] = 10/24$ and
\[
\left[
\begin{array}{ccc}
E[\alpha_1^2] & E[\alpha_1 \alpha_2] & E[\alpha_1 \alpha_3] \\
E[\alpha_2 \alpha_1] & E[\alpha_2^2] & E[\alpha_2 \alpha_3]\\
E[\alpha_3 \alpha_1] & E[\alpha_3 \alpha_2] & E[\alpha_3^2]
\end{array}
\right]
 =
 \frac{1}{20}
\left[
\begin{array}{ccc}
2 & 2 & 1\\
2 & 6 & 2\\
1 & 2 & 2
\end{array}
\right] ,
\]
whence $E [ \alpha_1 | x ] = 31/120$,  $E [ \alpha_2 | x] = 50/120$ and $E [ \alpha_3 | x ] = 39/120$. Thus $A_{21} / A_{12} = 86/46$ while
$B_{21} = m_2(x)/m_1(x) = 2$.

\bibliographystyle{chicago}
\bibliography{mix_BF_refs}

\end{document}